\documentclass[article]{agujournal2019}
\usepackage{url} 
\usepackage{lineno}
\usepackage{amsmath}
\usepackage[inline]{trackchanges} 
\usepackage{soul}
\usepackage{makecell}
\usepackage{multirow}

\usepackage[outline]{contour}

\journalname{JGR: Solid Earth}

\begin{document}

\title{Stationary phase analysis of ambient noise cross-correlations: Focusing on non-ballistic arrivals} 

\authors{Yunyue Elita Li\affil{1}, Feng Zhu\affil{1}, and Jizhong Yang\affil{2} }

\affiliation{1}{Department of Earth, Atmospheric, and Planetary Sciences, Purdue University, USA}
\affiliation{2}{State Key Laboratory of Marine Geology, Tongji University, China}

\correspondingauthor{Yunyue Elita Li}{elitali@purdue.edu}

\begin{keypoints}
\item We derive analytical stationary phase solutions for ambient noise cross-correlations with a focus on non-ballistic arrivals.
\item Non-ballistic arrivals in the stacked cross-correlation functions are not good approximations to the coda waves in the actual Green's function under the ambient noise condition. 
\item Changes in the non-ballistic arrivals cannot be uniquely attributed to changes in the medium or
changes in the noise source environment without additional constraints.
\end{keypoints}

\begin{abstract}
Stacked cross-correlation functions have become ubiquitous in the ambient seismic imaging and monitoring community as approximations to the Green's function between two receivers. While theoretical understanding of this approximation to the ballistic arrivals is well established, the equivalent analysis for the non-ballistic arrivals is alarmingly inadequate compared to the exponential growth of its applications. To provide a fundamental understanding of the cross-correlation functions beyond the ballistic arrivals, we derive analytical stationary phase solutions for ambient noise cross-correlations with a focus on non-ballistic arrivals. We establish the mathematical and corresponding physical conditions that drastically differentiate the non-ballistic arrivals in the stacked cross-correlation and the actual Green's functions. In ambient noise environments, the coda waves due to random medium scatterings of an impulsive source cannot be distinguished from the cross-talk artifacts due to overlapping random noise sources. Therefore, changes in the non-ballistic arrivals cannot be uniquely attributed to changes in the medium or changes in the noise source environment without additional constraints. The theoretical results demand that interpreting large-elapse-time arrivals in the stacked cross-correlation functions as coda waves for deterministic information about the propagation medium should be conducted only after the source influence is sufficiently ruled out. Once the source influence is eliminated, the stationary phase solutions for scattering waves provide a solid basis for extracting reliable scattering information from the noise correlation functions for higher-resolution imaging and monitoring.

\end{abstract}

\section*{Plain Language Summary}
Behind the ``magic'' of seismic interferometry that turns passive noise recording experiments into approximated active-seismic experiments are the mathematical operations of cross-correlation and averaging. While the magic works well for the waves that travel directly from one receiver to another, its validity for waves that have been scattered between the two receivers has not been thoroughly understood. To provide better clarity to this fundamental question, we derive mathematical and physical understandings of the averaged cross-correlation functions with a focus on their accuracy in approximating scattering events. We show that in ambient noise environments, the averaged cross-correlation functions are contaminated by source-induced cross-talk artifacts, making later arrivals in the cross-correlation functions indistinguishable from the random scatterings due to the impurities in the medium. We demonstrate a general equivalency between the later-time arrivals in the stacked cross-correlation functions and coda waves from impulsive sources does not exist. This theoretical study provides a solid foundation for evaluating and extracting reliable scattering information from the noise correlation functions for higher-resolution imaging and monitoring.

 \section{Introduction}
The operation of cross-correlation has become the foundation of ambient noise imaging and monitoring in the past decades. Many theoretical studies have shown that the ballistic wave Green's function can be obtained by the stacked cross-correlation functions of random fields recorded by two receivers. Theoretical understandings have been provided from the assumptions of equipartitioning of modal and propagating elastic vibrations \cite{lobkis2001emergence,sanchez2006retrieval}, stable time reversal of diffusive fields \cite{van2003green,wapenaar2004retrieving}, and the stationary phase analysis \cite{snieder2004extracting}. They are further verified by many laboratory studies \cite{weaver2001ultrasonics,derode2003estimate,malcolm2004extracting}. These studies focus on the ballistic component of Green's function, which is the solution of the wave equation for an impulsive point source in a \textit{background} (homogeneous or smooth) medium, represented by the ballistic (strongest energy) arrivals in the stacked cross-correlation function. 

In the field of reflection seismology, scattering/reflections from subsurface interfaces are of particular interest for imaging. \citeA{schuster2004interferometric} summarized the long history of conceptual and practical attempts to retrieve subsurface reflection seismograms from passive seismic energy first conjectured by \citeA{claerbout1968synthesis}. The focus is mostly on the interference of controlled sources that are widely available in the seismic exploration industry. 
Moreover, these practices do not provide sufficient theoretical understanding of the accuracy of such approximations.
\citeA{wapenaar2006green} presented a theoretical study where they claim crosscorrelations of full wavefields in arbitrary configurations produce the Green's function of the \textit{actual} medium, i.e., including scatterings from strong interfaces. However, the exactness of the proof is based on the following assumptions: 1) impulsive sources are placed at different locations, 2) the response of each source is measured separately, and 3) the availability of monopole- and dipole sources. When these conditions are not satisfied in practice, the equality between the \textit{actual} Green's function and the stacked cross-correlations becomes approximated. In the extreme case, where the sources are uncorrelated noise sources, this derivation reduces to the ones presented by \citeA{lobkis2001emergence}, \citeA{van2003green}, and \citeA{snieder2004extracting} under similar assumptions about the statistical properties of the sources.

In the field of earthquake seismology and acoustics, scattering effects from randomly distributed inhomogeneities have been studied to describe the property of the random medium \cite{knopoff1964scattering,aki1969analysis,aki1975origin,miles1960scattering} and to monitor subtle changes of the medium through coda wave interferometry \cite{snieder2006theory,pacheco2005localizing}. Coda waves are referred to as the scattered waves that come after the main (P-, S-, and surface wave) arrivals of an impulsive source. These scattered waves, as well as secondary microseism, are proven to be important sources of ambient seismic fields that enable the extraction of the Green's functions of the ballistic arrivals using cross-correlation. \citeA{sens2006passive} first hypothesized that the later part of the cross-correlation function corresponds to the scattered waves of the actual Green's function. This study showed empirically that coherent phases emerge in the stacked cross-correlation function after the ballistic arrival. These non-ballistic phases are then intuitively interpreted as the scattered waves between the two cross-correlated receivers, much the same way as coda waves from an impulsive source \cite{snieder2002coda,snieder2006theory}. Their study initiated a broad range of research utilizing the non-ballistic components of the cross-correlation (and the auto-correlation) function to monitor the changes of velocity in the scattered medium (e.g., \citeNP{brenguier2008towards,brenguier2008postseismic}), and subsequently map the changes in space (e.g., \citeNP{mao2022space}). 

Despite so many empirical successes of monitoring weak changes (on the order of 0.1\%) in the earth with ambient noise correlations, theoretical understanding and laboratory verification of their fundamental assumption, i.e., the cross-correlation functions produce the \textit{actual} Green's function, have not been established. To the contrary, \citeA{hadziioannou2009stability} showed via ultrasonic laboratory experiments that the correlation function from passive experiments is uncorrelated with the \textit{actual} Green's function from an active experiment. The accuracy of the measured velocity change depends heavily on the amount of repeatable ambient noise sources. Numerical studies \cite{clarke2011assessment,sheng2018nature} also pointed out such discrepancies between the stacked cross-correlation function (the ``empirical Green's function") and the  \textit{actual} Green's function. The increasing high-order applications of such monitoring methods and the lack of clarity necessitate fundamental understandings of the cross-correlation functions beyond the ballistic arrivals. 

In this paper, we present the stationary phase analysis of the non-ballistic arrivals in the stacked cross-correlation function. To avoid ambiguity, ``coda waves" are strictly referred to as later (scattered) arrivals from an \textbf{impulsive} source. The ``non-ballistic" arrivals are used to include both the precursory and the later arrivals in the \textbf{stack cross-correlation function}, compared to the ballistic arrival (i.e., the Green's function of the background medium). We establish the mathematical and corresponding physical conditions that drastically differentiate the non-ballistic arrivals in the stacked cross-correlation function and the actual coda waves. From the results of stationary phase analysis, we call for strong precautions when translating the characteristics of the non-ballistic arrivals measured in time to velocity changes in space. In particular, sensitivity kernels based on randomly scattering media for impulsive-source coda wave interferometry should not be applied blindly to the non-ballistic arrivals of the stacked cross-correlation function.

\section{Theory}

Inspired by \citeA{snieder2004extracting}, we perform stationary phase analysis of the cross-correlation function of random noise fields in 2D. We start from the most general assumption where the plane-wave noise sources are uncorrelated and randomly distributed in space and time. We will gradually relax this assumption to allow source correlations and generalize it to the case of multiple scattering in a randomly inhomogeneous medium.

\subsection{Cross-correlation of uncorrelated random sources}
\begin{figure}[h]
 \centering
 \includegraphics[width=0.7\textwidth]{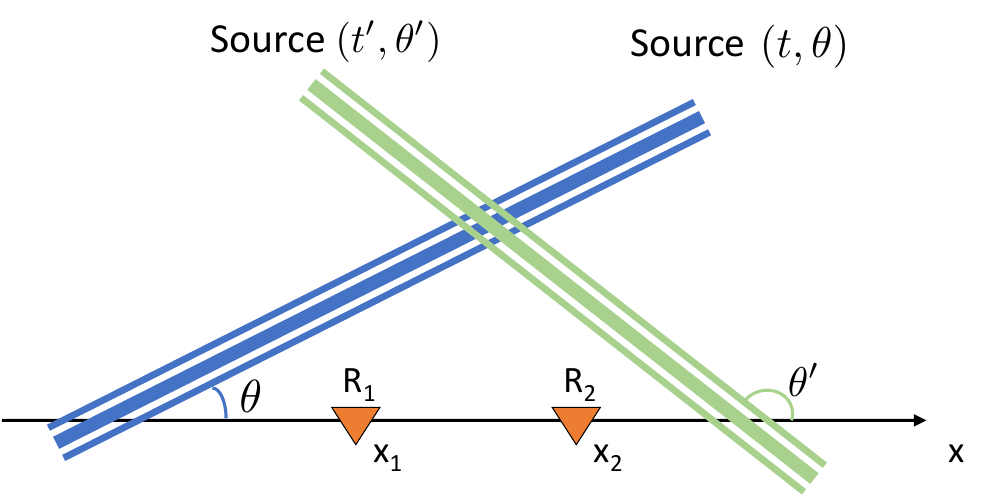}
 \caption{Schematic of the random plane-wave noise field and the acquisition geometry. }
 \label{fig:geometry}
 \end{figure}

Consider two receivers $R_1$ and $R_2$ that are deployed along a line defined by $\overrightarrow{x}$, and their respective locations are $x_1$ and $x_2$, as shown in Figure \ref{fig:geometry}. The medium is populated with plane-wave sources that are excited at time $t$ from $x=0$ with the wavefront angling at $\theta$ with respect to the receiver line $\overrightarrow{x}$. Denote the plane-wave source's frequency signature as $A(t,\theta;\omega)$, and we obtain a general expression of the overall noise field, which is a superposition of all random plane-wave sources
\begin{equation}\label{noisefield}
u(x,\omega)  = \sum_t \sum_\theta A(t,\theta;\omega) e^{-i\omega \left(t + \frac{x}{v / \sin \theta} \right) }, 
\end{equation}
where we further assume source excitation time $t$ and angle $\theta$ are random variables. The recordings at two receivers are then denoted as 
\begin{equation}\label{r1}
u_1(x_1,\omega)  = \sum_t \sum_\theta A(t,\theta;\omega) e^{-i\omega \left(t + \frac{x_1}{v / \sin \theta} \right) }, 
\end{equation}
and
\begin{equation}\label{r2}
u_2(x_2,\omega)  = \sum_t \sum_\theta A(t,\theta;\omega) e^{-i\omega \left(t + \frac{x_2}{v / \sin \theta} \right) }.
\end{equation}

The cross-correlation of these two recordings is computed by multiplication between the complex conjugation of one recording with the other in the frequency domain:
\begin{eqnarray}\label{cc}
u_1^*u_2(\omega) & = & \sum_t \sum_\theta A^*(t,\theta;\omega) e^{i\omega \left(t + \frac{x_1}{v / \sin \theta} \right) } \times  \sum_{t'} \sum_{\theta'} A(t',\theta';\omega) e^{-i\omega \left(t' + \frac{x_2}{v / \sin \theta'} \right) }, \nonumber \\
& = &  \sum_t \sum_\theta \sum_{t'} \sum_{\theta'} A^*(t,\theta;\omega) A(t',\theta';\omega) e^{-i\omega \left(t' -t - \left (\frac{x_1}{v / \sin \theta} - \frac{x_2}{v / \sin \theta'} \right) \right) }, 
\end{eqnarray}
where $^*$ denotes the complex conjugation. 
The phase of the cross-correlation function, 
\begin{equation}\label{phi_cc}
\phi_{cc} = t' - t + \left ( \frac{x_2}{v / \sin \theta'} - \frac{x_1}{v / \sin \theta}\right), 
\end{equation}
is a function of the random variables $t$, $t'$, $\theta$, and $\theta'$. Assuming all random plane-wave sources are {\bf uncorrelated}, the phase function behaves differently under the following scenarios:
\begin{itemize}
\item {\bf Scenario \#1: correlations of different sources} 

For recordings from different sources with $t \neq t'$ or $\theta \neq \theta'$, $\phi_{cc}$ is a random, variable function of the summation variables. This leads to the cross-correlation function vanishing, i.e., 
\begin{equation}\label{cc_diff_src}
u_1^*u_2(\omega) = \sum_t \sum_\theta \sum_{t'} \sum_{\theta'} A^*(t,\theta;\omega) A(t',\theta';\omega) e^{-i\omega \phi_{cc}} \rightarrow 0.
\end{equation}

\item {\bf Scenario \#2: correlations of the same source} 

For recordings from the same source with $t=t'$ and $\theta = \theta'$, the phase function becomes
\begin{equation}\label{phi_cc_ssrc}
\phi_{cc} = \frac{x_2- x_1}{v / \sin \theta}, 
\end{equation}
and the cross-correlation function becomes
\begin{equation}\label{cc_same_src}
u_1^*u_2(\omega) = \sum_t \sum_\theta |A(t,\theta;\omega)|^2 e^{-i\omega \phi_{cc}},
\end{equation}
whose phase function must be stationary with respect to $t$ and $\theta$ for the summation not to vanish. Hence, the condition requires 
\begin{eqnarray}\label{phi_cc_st}
\frac{\partial \phi_{cc}}{\partial \theta} & = & 0 \nonumber \\
\Rightarrow~~~ \frac{x_2- x_1}{v} \frac{\partial \sin \theta}{\partial \theta} & = & 0\nonumber \\
\Rightarrow~~~ \frac{x_2- x_1}{v} \cos \theta & = & 0, 
\end{eqnarray}
resulting in $\theta = \pm \frac{\pi}{2}$. Therefore, the stationary phases for the interference from the same sources are
\begin{equation}\label{phi_cc_ss}
\phi^{ss}_{cc} = \pm \frac{x_2- x_1}{v}, 
\end{equation}
and the cross-correlation function becomes
\begin{equation}\label{cc_same_src_st}
u_1^*u^{ss}_2(\omega) = \sum_t |A(t,\frac{\pi}{2};\omega)|^2 e^{-i\omega \frac{x_2-x_1}{v}} + \sum_t |A(t,-\frac{\pi}{2};\omega)|^2 e^{-i\omega \frac{x_1-x_2}{v}} ,
\end{equation}
where the first term accumulates all plane waves propagating along $\overrightarrow{x}$ from left to right, and the second term accumulates all plane waves propagating in the opposite direction. The superscript $ss$ stands for the same source. 

\end{itemize}

The analysis above reproduces the stationary phase analysis by \citeA{snieder2004extracting} in two dimensions. The assumptions about the random sources are critical. If all plane-wave sources are random and uncorrelated, the cross-correlation operation only retains the noise fields that are emitted from the same source, and propagate in parallel to the line determined by the two receivers. Therefore, in this most general (and most restrictive, at the same time) condition, stacking of the cross-correlation functions will eliminate all non-ballistic arrivals, as the number of stacks approaches the infinite limit. In practice, however, as the random and continuous noise sources always overlap in time and the number of sources cannot reach the infinite limit, the stacked cross-correlation functions are always contaminated by the correlations of recordings from different randomly, uncorrelated sources. These artifacts are randomly distributed, modulated by the squared average amplitude spectra of the noise sources, and may appear anywhere at any cross-correlation lags. 

This analysis, however, contradicts many empirical observations of stable non-ballistic arrivals in the cross-correlations functions (as first shown by \citeA{sens2006passive}). The stark discrepancy between the analysis and the practice leads us to consider conditions beyond the general random-source condition assumed in the last two scenarios: we should allow the sources to be correlated. Source correlations may be generated from two different origins: one from the noise source mechanism, such as period ocean waves striking the coastline, and the other from the correlations between primary and secondary sources, such as scattering waves generated by the propagation medium. We present the stationary phase analysis for a couple of special conditions and demonstrate that it is extremely challenging to distinguish these two origins of source correlations.

\subsection{Cross-correlations of time or angle correlated random sources} 

In this section, we consider sources may be correlated in \textit{either} time \textit{or} angle, and maintain the assumption that the time-angle correlation between sources is negligible. Under this scenario, the stationary condition requires
\begin{eqnarray}
\frac{\partial \phi_{cc}}{\partial t} = 0 & \contour{black}{$\Rightarrow$}  & \frac{\partial t'}{\partial t} - 1 = 0, \label{cccond1} \\
\frac{\partial \phi_{cc}}{\partial t'} = 0 & \contour{black}{$\Rightarrow$}  & 1- \frac{\partial t}{\partial t'}  = 0, \label{cccond2} \\
\frac{\partial \phi_{cc}}{\partial \theta} = 0 & \contour{black}{$\Rightarrow$}  & \frac{x_2\cos \theta'}{v} \frac{\partial \theta'}{\partial \theta} - \frac{x_1\cos \theta}{v} = 0, \label{cccond3} \\
\frac{\partial \phi_{cc}}{\partial \theta'} = 0 & \contour{black}{$\Rightarrow$}  & \frac{x_2\cos \theta'}{v}  - \frac{x_1 \cos \theta}{v} \frac{\partial \theta}{\partial \theta'} = 0. \label{cccond4}
\end{eqnarray}
The first two conditions \ref{cccond1} and \ref{cccond2} lead to a linear correlation between the source trigger times
\begin{equation}\label{cc-time-corr}
t' = t + t_c,
\end{equation}
where $t_c$ is an arbitrary time-delay function (can be positive or negative) independent of $t$. The last two conditions  \ref{cccond3} and \ref{cccond4} lead to deterministic conditions between the plane-wave source angles 
\begin{equation}\label{cc-angle-corr}
\theta' = \pm \theta \mathrm{~and~} \theta = \pm \frac{\pi}{2}.
\end{equation}
When the stationary phase conditions in Equations \ref{cc-time-corr} and \ref{cc-angle-corr} are satisfied, the cross-correlation phase function becomes: 
\begin{eqnarray}
  \phi^{cs}_{cc} =
    \begin{cases}
      t_c + \frac{x_2-x_1}{v}, & \mathrm{if}~ \theta=\frac{\pi}{2}, \theta'=\theta; \label{cc-sn1} \\
      t_c - \frac{x_2+x_1}{v}, & \mathrm{if}~ \theta=\frac{\pi}{2}, \theta'=-\theta; \label{cc-sn2}\\
      t_c - \frac{x_2-x_1}{v}, & \mathrm{if}~ \theta=-\frac{\pi}{2}, \theta'=\theta; \label{cc-sn3}\\
      t_c + \frac{x_2+x_1}{v}, & \mathrm{if}~ \theta=-\frac{\pi}{2}, \theta'=-\theta, \label{cc-sn4}\\
    \end{cases}      
    \label{cc-sphi-coda}
\end{eqnarray}
where the superscript $cs$ stands for correlated sources. These conditions require pairs of correlated sources. 
The stationary phase \ref{cc-sphi-coda} arises from the correlation between the first source recorded by $x_1$, and the corresponding second sources sending waves to $x_2$  with a time lag $t_c$. Both the first source and its correlated source should propagate along the line determined by the two receivers to satisfy the angle requirements. 

Figure \ref{fig:cc-corr-src} illustrates wavefield snapshots for the first two conditions in Equation \ref{cc-sphi-coda} where the same-direction propagation scenario is in (a) and the opposite-direction scenario is in (b). In both plots, the blue plane denotes the primary source, and the green plane denotes the correlated source. The last two cases in Equation \ref{cc-sphi-coda} correspond to the scenarios when the primary and correlated sources are placed at the mirror locations with respect to the center line between $x_1$ and $x_2$, i.e., the situations when the primary waves propagate from $x_2$ to $x_1$. Nonetheless, the cross-correlation functions are no longer symmetric, even if the primary source locations are uniformly distributed around the receivers. 

\subsubsection{Interpretation of the non-ballistic arrivals}

The results of the stationary phases can be explained by two different origins, as alluded to before. The first origin is from the correlation of the noise source function. Hence, $t_c$ is determined by the recurrence of the source events, such as the period of ocean waves hitting the coastlines and the average time interval between two motor vehicles running on the road. In these cases, {\it $t_c$ is a characteristic of the source}, and does not contain any information about the propagation medium. 

The second origin of the stationary phases is to consider the correlated source as a scattered wave (secondary source) of the first source at some interfaces in the propagation medium. We provide theoretical analyses for the scenarios of single and multiple scattering in the following discussion. While these specific geological conditions could generate data that fit the stacked cross-correlation functions, it is critical to understand the ambiguities between the source-induced correlations from the scattering-induced correlations. Further constraints are needed to uniquely attribute the physical origins of the non-ballistic arrivals.

\begin{itemize}

\vspace{0.1in}
\item {\bf The case of single scattering}
\vspace{0.1in}

In the case of single scattering that generates a correlated secondary source for each primary source, the blue and green planes in Figure \ref{fig:cc-corr-src} can be considered as the location of the source (such as the coastline) and the interface (such as a fault trace), respectively, at any given time. Assuming the distance between the primary and the secondary sources is $d$, we obtain $t_c$ as a function of the medium velocity $v$ \textit{between the two sources}, 
\[t_c = d/v.\] 
In this case, $t_c$ contains the information about the subsurface and are influenced by both $d$ and $v$. The non-ballistic arrivals in the cross-correlation function become 
\begin{equation}\label{cc_cs_st}
u_1^*u^{cs}_2(\omega) = \alpha\sum_t |A(t,\pm\frac{\pi}{2};\omega)|^2 e^{-i\omega \phi^{cs}_{cc}},
\end{equation}
where $\alpha$ is the scattering coefficient, which is the ratio between the secondary wave amplitude and the primary wave amplitude. When this happens, the stationary phases appear as distinct arrivals in the stacked cross-correlation function. However, a single measurement of $t_c$ results in an infinite number of possible geological conditions. The non-ballistic arrival could appear prior to (``precursory'') or after the ballistic arrival. In either case, these arrivals do not generally correspond to the physical scatterings of the ballistic arrivals in the cross-correlation functions. This is a fundamental difference between the stacked cross-correlation functions and the complex arrivals from an impulsive source, i.e., the actual Green's function. 

 \begin{figure}
 \includegraphics[width=\textwidth]{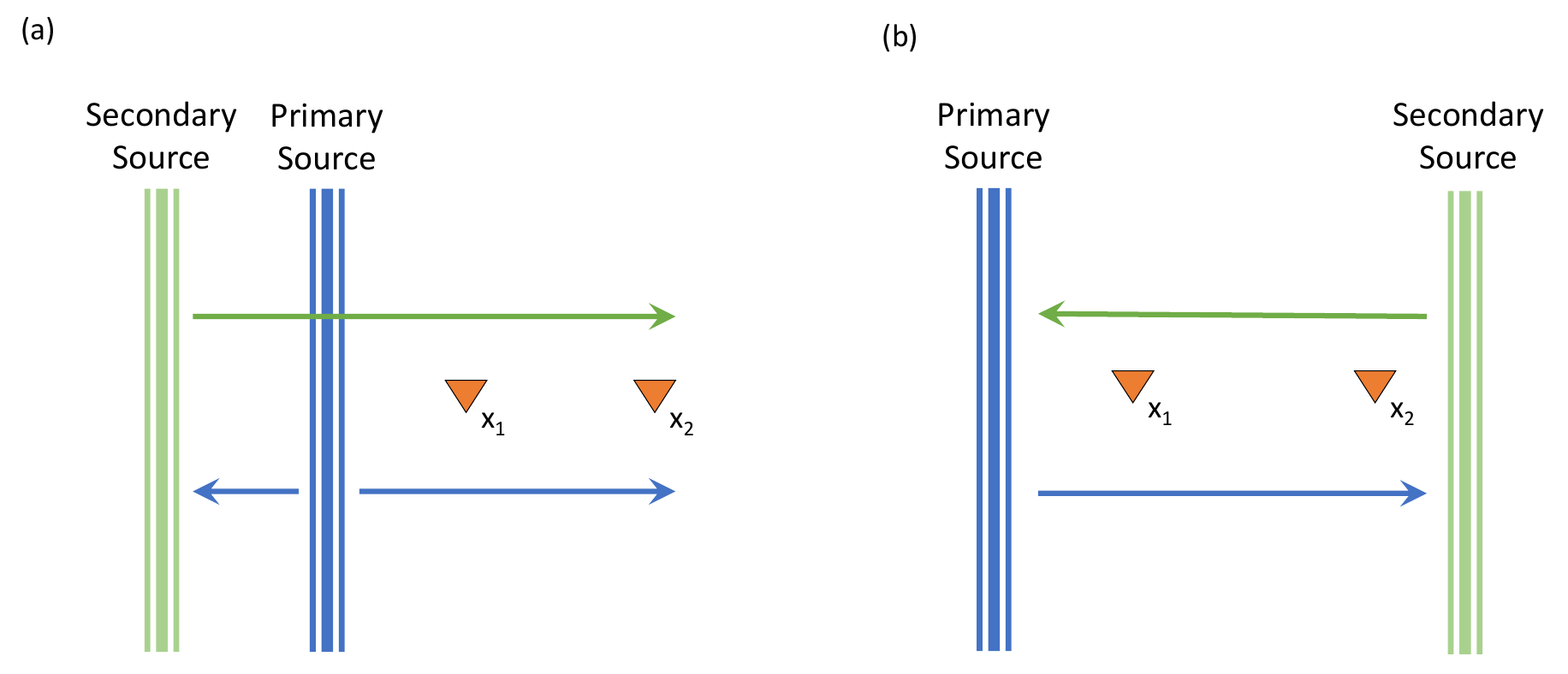}
 \caption{(a) Snapshots of the correlated wavefields for forward scattering. (b) Correlated source geometry for backward scattering. In both plots, the blue plane denotes the first source, and the green plane denotes the corresponding correlated source. The distance between these two sources is $d$. The stationary phases in the cross-correlation function correspond to the cross-correlation between the blue wavefront recorded by $x_1$ and the green wavefront recorded by $x_2$.}
 \label{fig:cc-corr-src}
 \end{figure}
 
\vspace{0.1in}
\item {\bf The case of multiple scatterings}
\vspace{0.1in}

 When the green and blue planes in Figure \ref{fig:cc-corr-src} represent two strong interfaces, the ambient noise fields can be scattered back and forth multiple times between these interfaces. Without losing generality and taking the case in Figure \ref{fig:cc-corr-src}(a) as an example, we obtain the correlated source time 
\begin{equation}\label{cc-time-corr-multsct}
t_c = (2n+1)  \frac{d}{v}, ~n =0, 1, \cdots, N, \cdots
\end{equation}
where $n$ is the number of multiple scattering between both interfaces.
The non-ballistic components of the stacked cross-correlation function then become
\begin{eqnarray}\label{cc_multsct_src}
u_1^*u^{cs}_2(\omega) & = & \sum_t |A(t,\pm \frac{\pi}{2};\omega)|^2  C \sum_{n=1}^{\infty} \frac{(\alpha_1\alpha_2)^n}{1-(\alpha_1\alpha_2)^{2n}} e^{\pm i \omega n \frac{2d}{v}},
\end{eqnarray}
where $\alpha_1$ and $\alpha_2$ represent the scattering coefficients of the two interfaces, respectively; and $C$ represent the constant phase-shift independent of the order of scattering. For example in the first scenario of Equation \ref{cc-sphi-coda}, \[C = e^{- i \omega \left ( \frac{d}{v} - \frac{x_2-x_1}{v} \right )}.\] Since the multiplication of the scattering coefficients is much smaller than one, i.e., $|\alpha_1 \alpha_2| \ll 1$, we further simplify the cross-correlation function as
\begin{equation}\label{cc_multsct_src_simp}
u_1^*u^{cs}_2(\omega) \approx \sum_t |A(t,\pm \frac{\pi}{2};\omega)|^2 C \sum_{n=1}^{\infty} (\alpha_1\alpha_2)^n e^{\pm i \omega n \frac{2d}{v}}.
\end{equation}
The cross-correlation function then represents an infinite time series with decaying amplitudes. The dominant frequency of the time series 
\begin{equation}\label{res-freq}
f_c = \frac{v}{2d} 
\end{equation}
is determined by the distance and the wave speed between the two strong interfaces. The non-ballistic arrivals in the stacked cross-correlation function are the recordings of the resonating waves between the two strong interfaces. As $|\alpha_1|$ and $|\alpha_2|$ can be both very small for weak scatters, the resonances are observed more often when one of the interfaces is the free surface, and when the wavelength of the propagating wave is on the same order as the distance between the two interfaces ($\lambda \sim d$).

\end{itemize}

\subsection{Cross-correlations of more strictly correlated random sources}

Wave propagation in an inhomogeneous medium naturally generates sources that are correlated not only in time, but also potentially in angle of propagation. In this section, we first ignore the source correlations, and present the analysis for two special scenarios where further dependence of the secondary source time and angle on the primary source time and angle is observed due to inhomogeneities of the propagation medium. From these analyses, we provide intuitive illustrations of the stationary phase zones for different arrivals in the stacked cross-correlation function. At the end of the subsection, we generalize the discussion to random media and discuss the various origins (source correlations, medium scatterings) of the non-ballistic arrivals and their relations to the actual Green's function. 

\subsubsection{Single point scatter in a homogeneous medium}

\begin{figure}[h]
    \centering
    \includegraphics[width=0.8\linewidth]{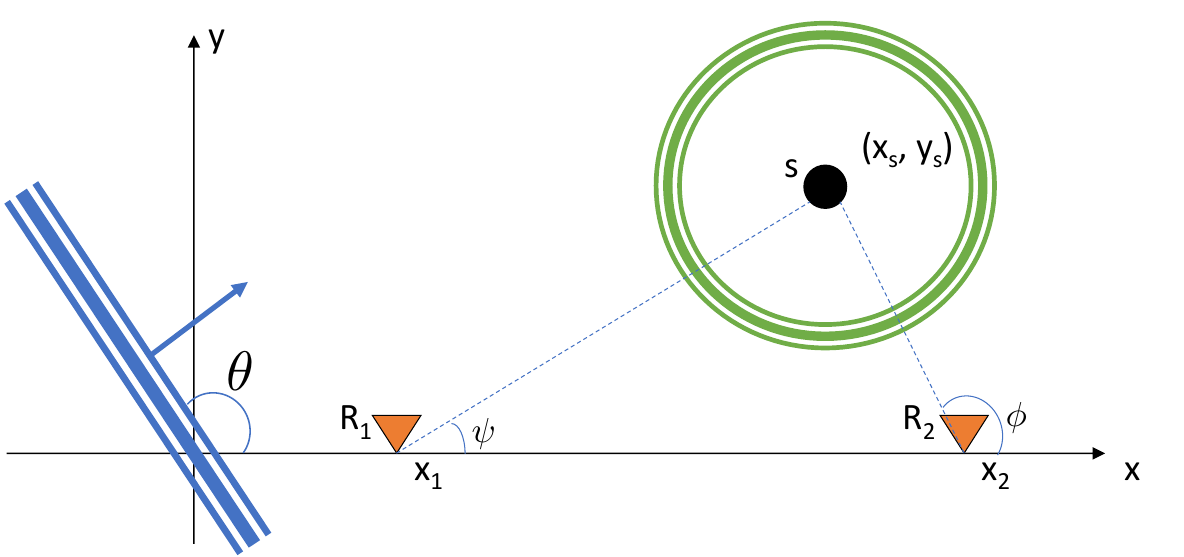}
    \caption{Sketch of the geometry for one point scatter in a homogeneous medium. }
    \label{fig:Point}
\end{figure}

We start from the simplest inhomogeneous scenario where a single point scatter is placed in the homogeneous medium with two receivers (Figure \ref{fig:Point}). Assuming a primary plane wave source 
\begin{equation}\label{prim-src}
    u_1 = e^{-i \omega (t - \frac{x \sin \theta}{v})}
\end{equation}
is set off at $t=0$, recording time of this primary wave at receiver $R_1$ located at $(x_1, 0)$ is
\begin{equation}\label{prim-r1-time}
    t_{R_1}^{u_1} = \frac{x_1 \sin\theta}{v},
\end{equation}
and the arrival time of this primary wave at the scattering location $(x_s, y_s)$ is
\begin{equation}\label{prim-sct-time}
    t_{s}^{u_1} = \frac{x_s \sin\theta}{v} + \frac{y_s \sin(\theta-\frac{\pi}{2})}{v}. 
\end{equation}

As soon as the primary source reaches the scattering location, the point scatter acts as a secondary source, sending out a circular wave 
\begin{equation}\label{sct-src}
    u_2 = \alpha_{ps} e^{-i \omega (t - t_{s}^{u_1} - \frac{d_s}{v})},
\end{equation}
where $d_s = \sqrt{(x-x_s)^2 + (y-y_s)^2}$ is the distance to the secondary source and $\alpha_{ps}$ is the scattering coefficients of the point scatter. Therefore, the secondary source is observed at receiver $R_2$ at $(x_2, 0)$ at the arrival time of 
\begin{equation}\label{sct-R2-time}
    t_{R_2}^{u_2} = t_{s}^{u_1} + \frac{d_{s,R_2}}{v},
\end{equation}
where $d_{s,R_2} = \sqrt{(x_2-x_s)^2 + y_s^2}$. \textit{Assuming $R_1$ only records $u_1$, and $R_2$ only records $u_2$}, the phase function in the cross-correlation of the two recordings is
\begin{eqnarray}\label{phi-PS}
    \phi_{cc}^{u_1,u_2} & = & t_{R_2}^{u_2} - t_{R_1}^{u_1} \nonumber \\
    & = & \frac{(x_s-x_1)\sin\theta}{v} - \frac{y_s\cos \theta}{v} + \frac{d_{s,R_2}}{v}.
\end{eqnarray}
The stationary phase condition $\frac{\partial \phi_{cc}^{u_1,u_2}}{\partial \theta} = 0 $ leads to the following constraints
\begin{eqnarray}\label{stn-phi-PS}
   & & (x_s-x_1)\sin\theta + y_s\sin \theta = 0, \\
   & \Rightarrow & \tan \theta = - \frac{x_s-x_1}{y_s}.
\end{eqnarray}
From simple trigonometry as sketched out in Figure \ref{fig:Point}, we know the angle between the line connecting $R_1$ and the scatter and the $x$-axis $\psi$ follows
\begin{equation}\label{R1-PS-line}
    \cot \psi = \frac{x_s-x_1}{y_s} = \cot(\theta - \frac{\pi}{2}) \Rightarrow \theta = \psi + \frac{\pi}{2}.
\end{equation}
Hence, the stationary phase in the cross-correlation arrives at  
\begin{equation}\label{Single-GF}
    \phi_{cc,stn}^{u_1,u_2} = \frac{d_{s,R_1}}{v} + \frac{d_{s,R_2}}{v},
\end{equation}
which coincides with the actual arrival of the scattered wave when an impulsive source is set off at $R_1$. Similarly, if we move the reference coordinate frame to the right-hand-side of $x_2$, and allow the plane waves to propagate to the negative $\overrightarrow{x}$ direction, we will obtain another stationary phase arrival 
\begin{equation}\label{Single-GF-negx}
    \phi_{cc,stn}^{u_1,u_2} = -(\frac{d_{s,R_1}}{v} + \frac{d_{s,R_2}}{v}),
\end{equation}
with the wavefront angle $\theta = \phi - \frac{\pi}{2}$.

From the analysis, we show that wave propagation in inhomogeneous media automatically generates correlated sources. In the case of point scattering, the secondary source (the scattered wave) time is determined by the angle of the primary source, while the secondary source angle is independent of the angle or time of the primary source.

\begin{figure}[ht]
    \centering
    \includegraphics[width=\linewidth]{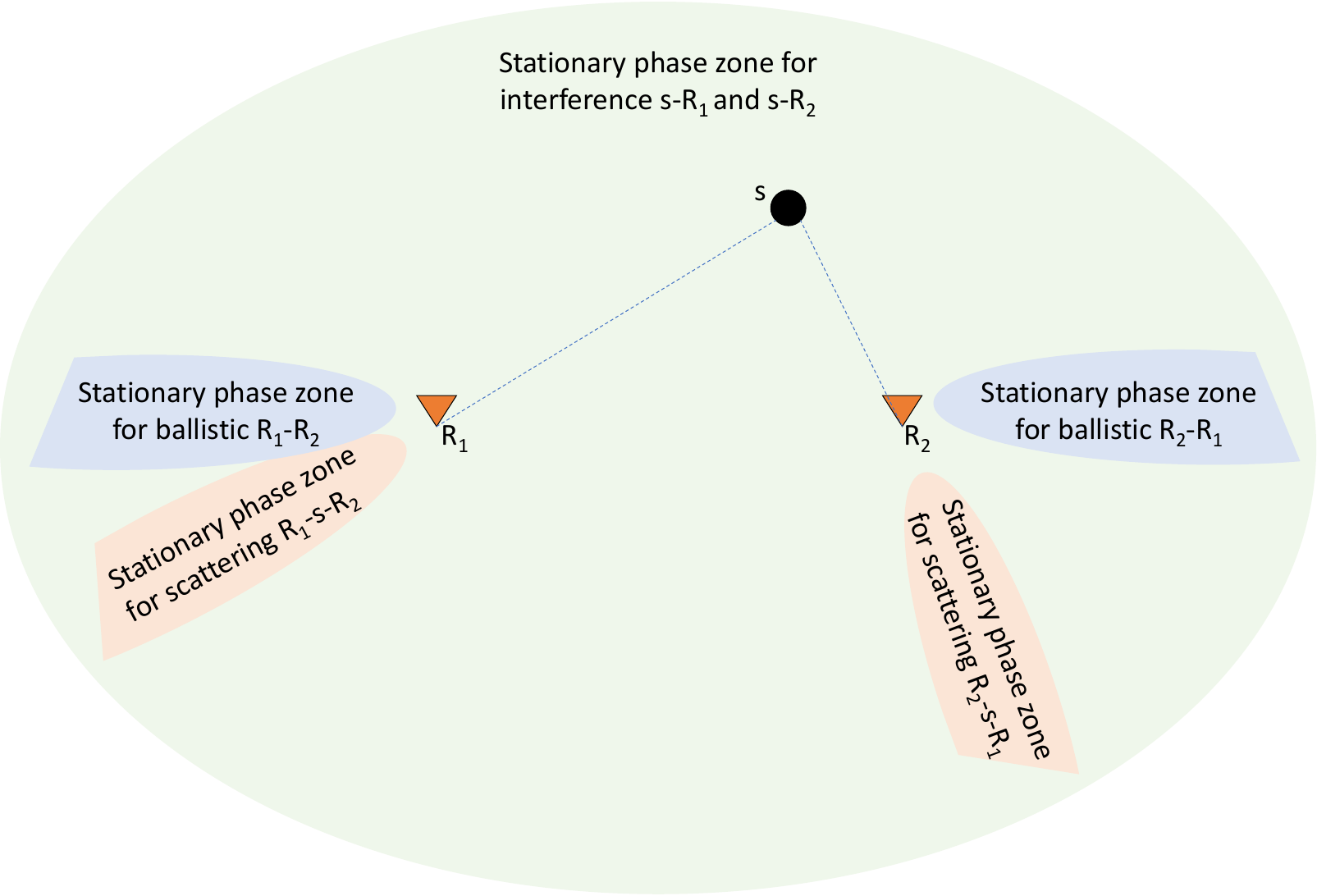}
    \caption{Illustration of the stationary phase zones for different arrivals in the stacked cross-correlation function. }
    \label{fig:Pt_SPZ}
\end{figure}

Equation \ref{Single-GF} and \ref{Single-GF-negx} suggest that the single-scattering phases as part of the actual Green's function between $R_1$ and $R_2$ are preserved when primary sources are excited at specific angles. However, this is achieved at a strong condition that  
\textit{$R_1$ only records the primary source, and $R_2$ only records the secondary source}. In reality, recordings from primary and secondary sources cannot be separated in ambient noise. Therefore, additional phases will also be stacked constructively. For example, another stationary arrival in the cross-correlation function is the interference of the scattered wave $u_2$ recorded at both receivers:
\begin{equation}\label{Single-GF-XT}
    \phi_{cc,stn}^{u_2,u_2} = \frac{d_{s,R_2}}{v} - \frac{d_{s,R_1}}{v}.
\end{equation}
Depending on the strength of the scattering, individual amplitudes of such phases ($O(\alpha_{ps}^2)$) may not be comparable to the amplitudes of $\phi_{cc,stn}^{u_1,u_2}$ ($O(\alpha_{ps})$). However, its stationarity holds for all random primary sources, making it a non-negligible contribution in the stacked cross-correlation function. Since $d_{s,R_2} - d_{s,R_1}$ is always smaller than $d_{R_1,R_2}$, the cross-talk phase always arrives earlier than the ballistic waves, contributing to the ``precursory" arrivals in the stacked cross-correlation function.  



Figure \ref{fig:Pt_SPZ} illustrates the stationary phase zones for different arrivals in the stacked cross-correlation function. When primary noise sources fall in these zones, the respective arrivals will be stacked constructively. The light blue regions denote the stationary phase zones for the ballistic arrivals between the two receivers, while the light brown regions are for the scattering arrivals. Compared to more restricted zones of these physical phases, the stationary phase zone for the interference between the scattering wave $s-R_1$ and $s-R_2$ covers the whole 2D domain. 

\begin{figure}[h]
    \centering
    \includegraphics[width=\columnwidth]{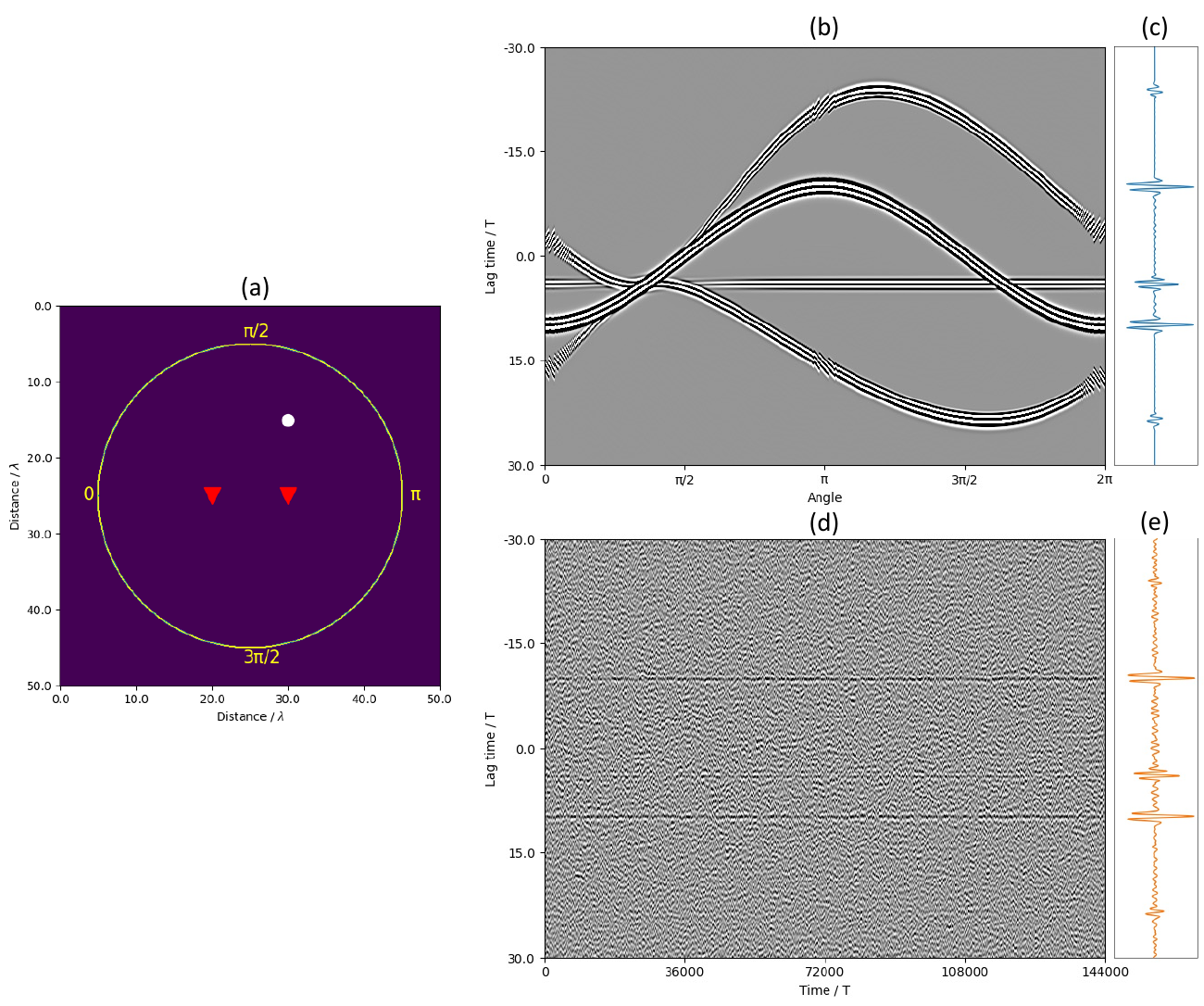}
    \caption{(a) Source and receiver geometry of the numerical simulation. The sources are located on a full circle around the two receivers, denoted by the red triangles. The medium is homogeneous except for a point scatter denoted by the white dot. (b) Cross-correlation functions for individual impulsive sources from $0$ to $2\pi$. (c) The stacked cross-correlation function of (b). (d) Cross-correlation functions for random, uncorrelated, and continuous sources recorded at different times. (e) The stacked cross-correlation function of (d).}
    \label{fig:Pt_XC}
\end{figure}

We further demonstrate the theoretical results using numerical simulations. Figure \ref{fig:Pt_XC}(a) shows the geometry of the numerical simulation, where sources and receivers are placed in a homogeneous medium with a single scattering point denoted by the white dot. The sources are distributed uniformly around the two receivers. When these sources fire impulsive energy and are recorded individually by the two receivers, the cross-correlation functions for each individual source are plotted in Figure \ref{fig:Pt_XC}(b) and their stack is plotted in Figure \ref{fig:Pt_XC}(c). Evidently, besides the stationary phases at $0$ and $\pi$ that correspond to the ballistic arrivals between the two receivers, there are three additional stationary phases in the cross-correlation function. The two symmetric phases correspond to the scattering wave paths $R_1-s-R_2$ and $R_2-s-R_1$ with the stationary phase angle determined by $R_1\rightarrow S$ and $R_2\rightarrow s$, respectively. Due to narrow stationary phase zones and a small scattering coefficient, these phases are much weaker compared to the ballistic phases. The third phase, arriving prior to the ballistic arrivals with a positive time lag, is stationary with respect to all sources. This is the interference between the scattering wave $s-R_1$ and $s-R_2$. While the amplitude of each individual interference is small, the full-range stationarity significantly increases its amplitude in the stacked cross-correlation function. This example demonstrates that even under idealistic source conditions, the stacked cross-correlation function can significantly differ from the actual Green's function. 

Figure \ref{fig:Pt_XC}(d) and (e) shows the cross-correlation functions and their stacks when the sources fire random, uncorrelated, but continuous energies with the same amplitude spectrum as the impulsive sources in Figure \ref{fig:Pt_XC}(b) and (c). In this case, both receivers record overlapping sources that cannot be separated. The cross-correlation functions in (d) are shown in the order of their recording time, within which all sources have possibly sent out energies from all angles. When sources are overlapped in the recordings, source cross-talks overwhelm the individual cross-correlation function. While the stationary phase components are enhanced after stacking and the source cross-talk artifacts are significantly reduced, the artifacts cannot be fully removed. These artifacts dramatically reduce the signal-to-noise ratio (SNR) for the scattering phases $R_1-s-R_2$ and $R_2-s-R_1$. When the scattering coefficient of the point scatter is smaller, or the number of random source stacks is reduced, these scattering phases can be easily buried below the cross-talk artifacts.

\subsubsection{Planar interface in a homogeneous medium}

\begin{figure}[h]
    \centering
    \includegraphics[width=0.8\linewidth]{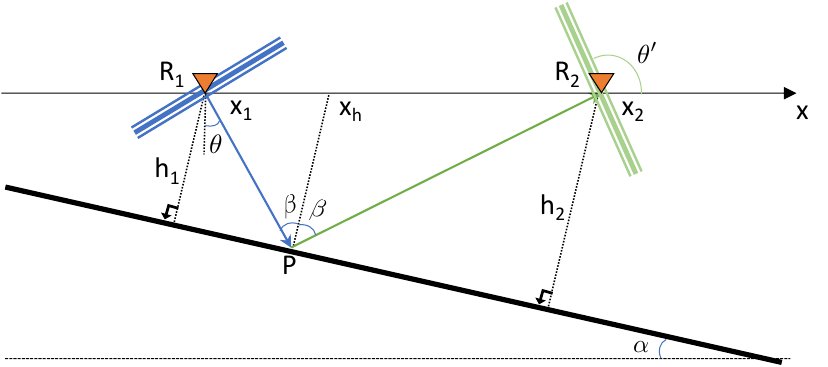}
    \caption{Illustration of the stationary phase zones for different arrivals in the stacked cross-correlation function. }
    \label{fig:PlaneS}
\end{figure}

The second special case of an inhomogeneous medium concerns a planar interface in a homogeneous background. We further demand the interface does not intercept the section determined by $R_1$ and $R_2$. This is equivalent to requiring all (primary and secondary) sources outside of the support of the receiver array $R_1$-$R_2$. The case when they intercept is very different from the following discussion. The geometry of the special case is illustrated in Figure \ref{fig:PlaneS}, where two receivers are denoted by the yellow triangles, and the interface is denoted by the thick black line. 

Given the planar reflector $y = \tan (\alpha) x + b$, and the two receiver locations $(x_1,0)$ and $(x_2,0)$, Snell's law determines the specular incident ray (blue arrow) and the specular reflection ray (green arrow) that connect $R_1$ with $R_2$, if a source is set out at $R_1$. The incident and the reflection angles with respect to the normal of the reflector are the same:
\begin{equation} \label{inciangle}
    \beta = \theta + \alpha, 
\end{equation}
where $\theta$ is the angle of the primary source $u_1$, 
\begin{equation}
    u_1 = e^{-i\omega(t-\frac{x\sin\theta}{v})}.
\end{equation}
The angle of the secondary source (specularly reflected wave) $u_2$ is determined by the angle of the incident wave and the angle of the reflector
\begin{equation} \label{refangle}
    \theta' = \pi - (\theta + 2\alpha).
\end{equation}
By observing the geometry, we obtain the following system of equations where $h_1$ and $h_2$ are known distances from the receivers $R_1$ and $R_2$ to the planar reflector, respectively: 
\begin{eqnarray} \label{angle-system}
    \frac{x_h-x_1}{\sin \beta} & = & \frac{h_1/\cos\beta}{\sin(\pi/2 - \theta)}, \nonumber \\ 
    \frac{x_2-x_h}{\sin \beta} & = & \frac{h_2/\cos\beta}{\sin(\theta' - \pi/2)}. 
\end{eqnarray}
Substituting Equations \ref{inciangle} and \ref{refangle} into the system above, we are left with two unknowns $x_h$ and $\theta$. Both $x_h$ and $\theta$ are uniquely solved, as long as the reflector does not intersect the section determined by $R_1$ and $R_2$. Consequently, the incident angle $\theta$ and the specular reflection point $P$ are uniquely determined for fixed $R_1$, $R_2$, and planar reflector geometry. 

The secondary source $u_2$ as the specularly reflected wave of the primary source $u_1$ is then 
\begin{equation}
    u_2 = \alpha_{pl} e^{-i\omega(t-t^{u_1}_{R_1}- \frac{d_{R_1,P,R_2}}{v})},
\end{equation}
where $\alpha_{pl}$ is the reflection coefficient of the planar reflector and $d_{R_1-P-R_2}$ denotes the summation of the distances between $R_1$ and $P$ and between $P$ and $R_2$. The arrival time of the secondary source at $R_2$ is then determined 
\begin{equation}\label{Plane-u2-time}
    t^{u_2}_{R_2} = t^{u_1}_{R_1} + \frac{h_1/\cos\beta}{v} + \frac{h_2/\cos\beta}{v}. 
\end{equation}
It is now trivial to show that the stationary phase resulted from the correlation between $u_1$ and $u_2$ is
\begin{eqnarray} \label{Single-PlaneS}
    \phi_{cc,stn}^{u_1,u_2} & = & \frac{h_1/\cos\beta}{v} + \frac{h_2/\cos\beta}{v}, \nonumber \\
    & = & \frac {h_1} {\cos(\theta + \alpha)v} + \frac{h_2}{\cos(\theta + \alpha)v}.
\end{eqnarray}
From Equations \ref{refangle} and \ref{Plane-u2-time}, it is obvious that both the angle and time of the secondary source are fully determined by the time and angle of the primary source. Similarly, we will obtain the negative lag stationary phase arrival if we allow the primary source first pass through $R_2$, reflect off the plane, and be recorded by $R_2$.

\begin{figure}[h]
    \centering
    \includegraphics[width=\linewidth]{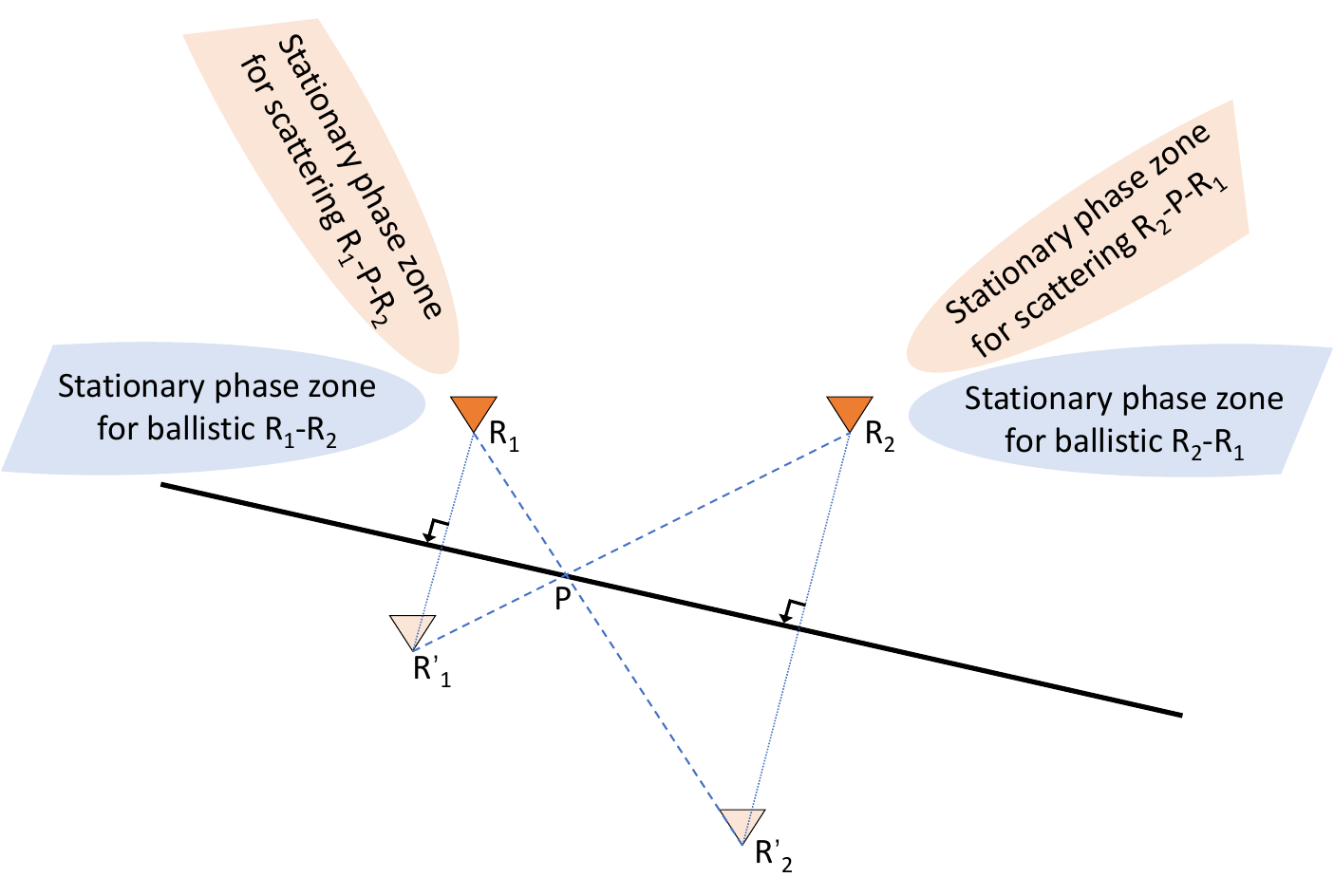}
    \caption{Illustration of the stationary phase zones for different arrivals in the stacked cross-correlation function. }
    \label{fig:Pl_SPZ}
\end{figure}

The above analysis shows that stationary phase arrivals are preserved in the stacked cross-correlation function for the specular reflection in the actual Green's function. Intuitively, we can identify the stationary phase zones for the specular reflections using the ``mirror images" of the receivers with respect to the planar reflector. In Figure \ref{fig:Pl_SPZ}, the light shaded triangles $R'_1$, and $R'_2$  denote the mirror images of $R_1$ and $R_2$, respectively. The stationary phase zones of the specular reflections (the light brown regions) are then readily identified by the line determined by the actual receiver $R_1$ ($R_2$) and the mirror image of the other receiver $R'_2$ ($R'_1$).  

Thanks to the strict angle correlation between the primary and secondary waves determined by Shell's law, there are no additional artifacts caused by the scattering as in the case of the point scatter. Consequently, when the primary noise source correlation can be fully ruled out, the stacked cross-correlation function fully recovers the direct waves and the first-order scattering wave. This is a reason why seismic interferometry has been successfully applied in controlled source experiments where each shot is recorded individually \cite{schuster2004interferometric,wapenaar2010tutorialP1,wapenaar2010tutorialP2}. When there is more than one reflector in the subsurface, cross-talks generated by the scatterings from different reflectors contaminate the stacked cross-correlation function and can be removed by various interferometric inversion schemes (e.g., \citeNP{zhu2022extension}).

\begin{figure}[ht]
    \centering
    \includegraphics[width=\linewidth]{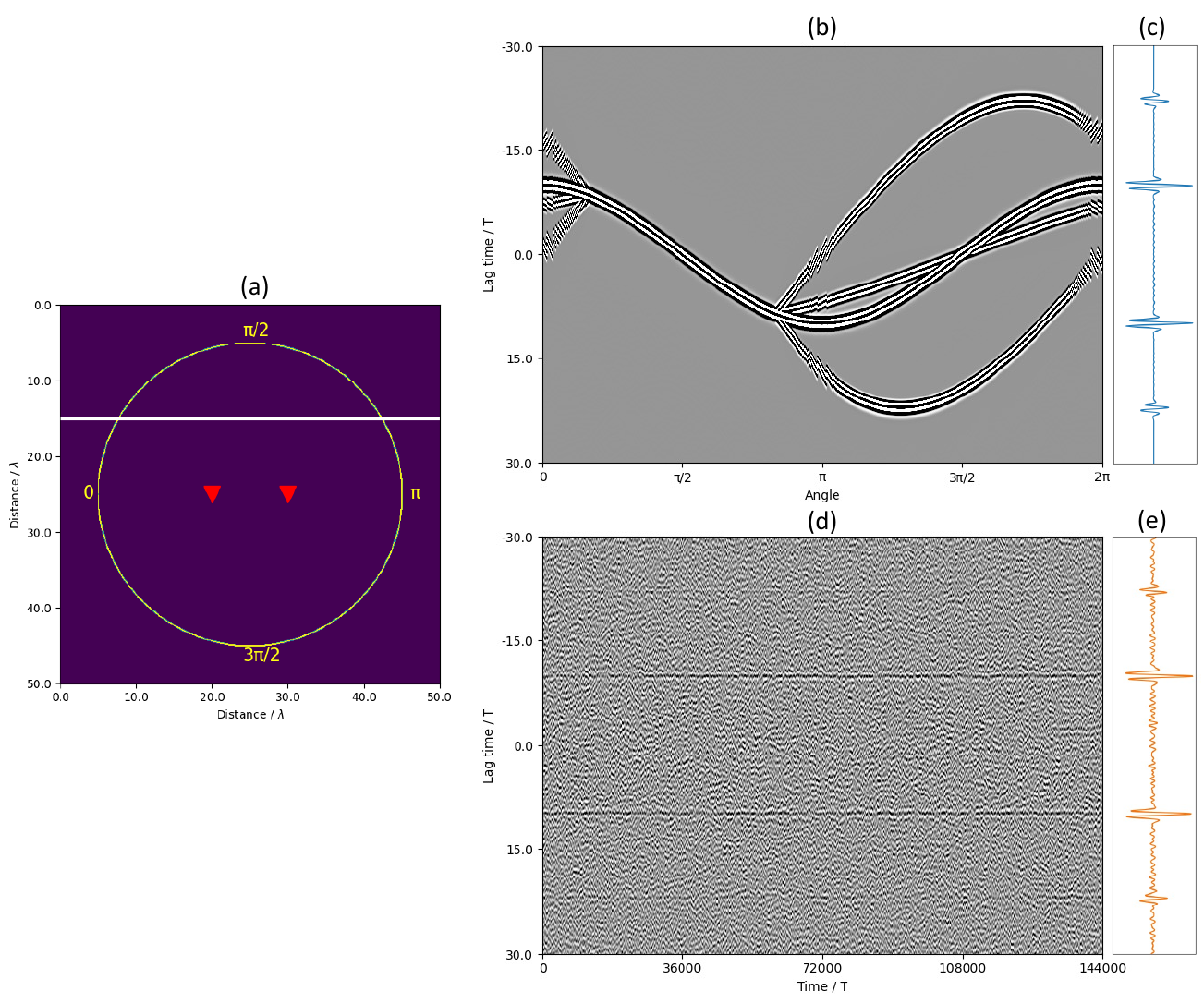}
    \caption{Subplots are in the same order as in Figure \ref{fig:Pt_XC}. The only difference is in the medium for simulation, where a single planar reflector (denoted by the white line) is present in a homogeneous background medium.}
    \label{fig:Pl_XC}
\end{figure}

We perform the same numerical simulation as for the case of a point scatter and plot the results in Figure \ref{fig:Pl_XC}. As clearly shown in Figure \ref{fig:Pl_XC}(b) and (c), the specular reflections are faithfully reconstructed at the corresponding stationary phase angles when impulsive sources are recorded individually. This nearly perfect reconstruction is contaminated by source cross-talk artifacts when both receivers record overlapping random sources. Similar to the case of point scattering, specular reflection phases in the stacked cross-correlation function can be buried under the source cross-talk artifacts if the planar interface is not strong enough or the stacking power is insufficient. 

These theoretical and numerical results also explain why reconstructing scattered body waves is extremely challenging in the ambient noise environment on land, where uncontrolled, uncorrelated, and random sources generate much stronger surface waves than body waves. Cross-talk artifacts from the correlations of the surface wave sources overwhelm the amplitudes of the scattered body waves in the non-ballistic arrivals, making them even less likely to be observed than the scattered surface waves from the same interface. 

\subsubsection{Scattering in random media}

The non-ballistic arrivals in the stacked cross-correlation functions are often discussed under a general context of scattering in random media. Conceptually, we construct the random media using the superposition of individual point scatters. When the point scatters are randomly distributed around the receivers, the stationary phase zones to recover the Green's function components of the first-order scattering cover the whole 2D domain, rotating with the lines determined by $R_1-s$ and $R_2-s$ (Figure \ref{fig:Pt_SPZ}). Meanwhile, these scatters also generate stationary phases that are constructively stacked before the ballistic arrivals, as discussed in Equation \ref{Single-GF-XT}. When there are no overlapping sources for each cross-correlation time window, the stacked cross-correlation function could be a good approximation to the actual Green's function in the random media. However, this approximate breaks down in practice, because the cross-talks from the overlapping noise sources cannot be fully eliminated, even when the noise sources are uniformly distributed around the receivers and the scatters. 

\begin{figure}[ht]
        \centering
        \includegraphics[width=\columnwidth]{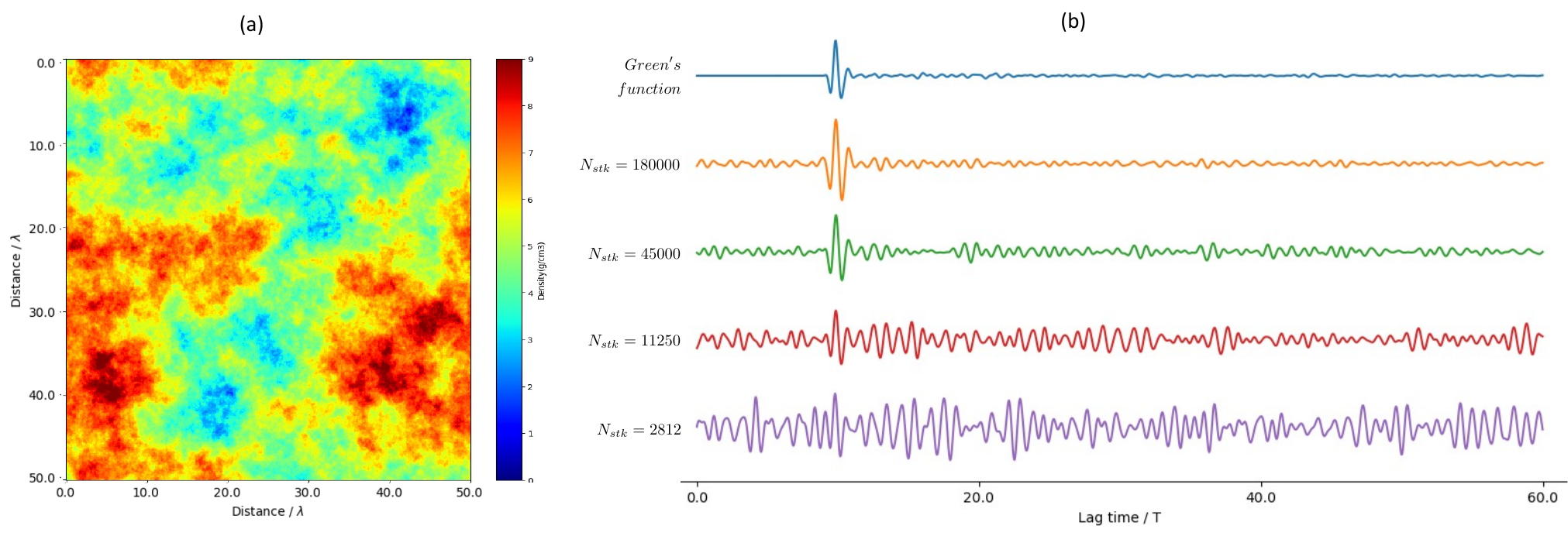}
        \caption{(a) Random density model for the numerical simulation. (b) Comparison between the actual Green's function and the normalized XCFs after $N_{stk}=\{2812, 11250, 45000, 180000\}$ stacks. }
        \label{fig:stackpower}
    \end{figure}

Figure \ref{fig:stackpower}(a) shows the random density model we use to perform the wave simulation in a random scattering medium. We intentionally create a strong scattering regime by varying the densities more than 100\% around their mean value. We use the same homogeneous velocity model to ensure the ballistic arrivals occur at the same time as the homogeneous case. We simulate recordings at the two receiver locations from random, uncorrelated, and continuous sources. Figure \ref{fig:stackpower}(b) shows the comparison between the actual Green's function and the normalized stacked cross-correlation functions after $N=\{2812, 11250, 45000, 180000\}$ stacks. We observe $\sqrt{N}$-rate improvement of the SNR as expected. The ballistic arrival gradually stands out from the source cross-talks. However, the coda waves in the actual Green's function are so weak that even at the maximum stacking power $N = 180000$, their SNR is still less than 1:1, resulting in no similarity between the non-ballistic arrivals in the stacked cross-correlation functions and the actual coda waves due to random scattering.

In Table \ref{tab:OriginXCF}, we summarize different arrivals and artifacts that may be observed in the cross-correlation function. We compare their origins, arrival times, and amplitudes with respect to the ballistic arrivals and their relations to the actual Green's function components. We limit our discussion to the first-order scattering of the primary noise sources. When the primary noise sources are uniformly distributed, the higher-order scatterings have much lower amplitudes and will not stand out from the cross-talk artifacts. 

We use $\bar{A}$ to represent the average amplitude spectrum of the primary noise sources. Ignoring all higher-order correlations, the amplitude of the ballistic arrival in the stacked cross-correlation function is proportional to the number of random sources in the stationary phase zone for the direct wave ($N_{dw}^{spz}$). Similarly, amplitudes of all other stationary phase arrivals are proportional to the number of random sources in the respective stationary phase zones. However, their amplitudes are further reduced proportional to the scattering coefficients of the medium heterogeneities. 

Amplitudes of the cross-talk artifacts are determined by two factors: they are proportional to the number of overlapping sources $N_{ol}$ in the cross-correlation time window and inversely proportional to the square root of the number of stacks $N_{stk}$. This increase in SNR relies on the random nature of the noise sources. If noise sources become stable over the stacking time, the cross-talk artifacts will not be reduced according to the inverse of the square-root law and spurious arrivals will be maintained.

\begin{table}[ht]
    \centering
    \begin{tabular}{|c|c|c|c|c|}
        \hline
        \makecell{Math\\ Origin} & \makecell{Physics Origin} & \makecell{Arrival time} & \makecell{Amplitude} & \makecell{GF \\ Comp.} \\
        \hline
        {\makecell{Same\\ Sources}} & \makecell{Same \\ Primary Sources} & Ballistic & $N_{dw}^{spz}\bar{A}^2$ & Y \\
        \hline
        \multirow{2}{*}{\makecell{Correlated\\ Sources}}  & \makecell{Primary and its \\ point scattering} & Aft & $ N_{ps}^{spz} \alpha_{ps}\bar{A}^2$ & Y \\
        \cline{2-5}
         & \makecell{Primary and its \\ planar scattering} & Aft & $ N_{pl}^{spz}\alpha_{pl}\bar{A}^2$ & Y  \\
        \hline
        \multirow{2}{*}{\makecell{Different\\ Sources}} & \makecell{Overlapping \\ Primary Sources} & Pre \& Aft & $N_{ol}\bar{A}^2/\sqrt{N_{stk}}$ & N \\
        \cline{2-5}
         & \makecell{Primary and other \\ source's scattering} & Pre \& Aft & $N_{ol}\alpha\bar{A}^2/\sqrt{N_{stk}}$ & N \\
        \hline
    \end{tabular}
    \caption{Origins, arrival times, and amplitudes of arrivals in the stacked XCF.}
    \label{tab:OriginXCF}
\end{table}

Comparing the amplitudes of the scattered waves in the stacked cross-correlation function with those of the artifacts, we stress the importance of the availability of the noise sources in the respective stationary phase zones for the scatterings. In general, stationary phase amplitudes $N_{ps,pl}^{stn}\alpha_{ps,pl}$ have to be significantly larger than $N_{ol}/\sqrt{N_{stk}}$ to ensure reliable observations of the primary scattered waves. As the scattering coefficients of a point scatter $\alpha_{ps}$ can be an order of magnitude smaller than those of planar interfaces $\alpha_{pl}$, scatterings off planar geological interfaces are more likely to be observed in the stacked cross-correlation function.
Optimizing scattering wave reconstruction requires maximizing the number of noise sources in the stationary phase zones, minimizing the number of overlapping noise sources within the cross-correlation window, and maximizing the stacking number of random noise sources. Unless the geological condition is extremely favorable, higher-order scatterings may be completely buried under the source cross-talk artifacts.

\section{Discussion}

\subsection{Key insights from the theoretical analysis}

The analysis results show that the non-ballistic arrivals in the stacked cross-correlation functions could carry drastically different physical meanings compared to the coda waves from impulsive sources, such as earthquakes \cite{knopoff1964scattering,aki1969analysis,aki1975origin} or controlled seismic sources \cite{schuster2004interferometric}. Using the following bullet points, we stress the key insights from the analysis. 
\begin{itemize}
    \item Stacked cross-correlation functions are {\bf NOT} the same as Green's function, even when the random sources are uniformly distributed around the receivers. 
    \item In practice, the coda waves due to random scatterings of an impulsive source (controlled by the medium properties) cannot be distinguished from the cross-talk artifacts (controlled by the random source properties) in the non-ballistic arrivals of the stacked cross-correlation function. 
    \item When stable non-ballistic arrivals appear in the stacked cross-correlation functions, they could be due to correlations of the environmental sources, or scatterings of sources in different stationary phase zones.
    \item The stationary phase zones to properly reconstruct scattering waves in the stacked cross-correlation function (non-ballistic arrivals) are markedly different from those to properly reconstruct the direct waves (ballistic arrivals). 

\end{itemize}

These insights demand extreme caution when non-ballistic arrivals in the stacked cross-correlation functions are interpreted. In practice, changes in the non-ballistic arrivals cannot be uniquely attributed to changes in the medium or changes in the noise source environment without additional constraints. Interpreting large-elapse-time arrivals in the stacked cross-correlation functions as coda waves for deterministic information about the propagation medium should be conducted only after the source influence is sufficiently ruled out.  

\subsection{On the ambiguities between the source and medium effects}
In this section, we use a couple of numerical examples to highlight the theoretical results on the general ambiguity between the source interference and the medium scattering effects. Artifacts in the stacked cross-correlation function may arise from many aspects in field data. To avoid further complications, we assume the primary noise sources are uniformly distributed around the two receivers and outside the section defined by the two receivers. We stacked the cross-correlation functions with sufficient sources and recording duration, such that the non-ballistic arrivals are stable and their amplitudes cannot be further decreased with respect to the ballistic arrival. We assume no intrinsic attenuation during wave propagation. 

\begin{figure}[ht]
    \centering
    \includegraphics[width=\linewidth]{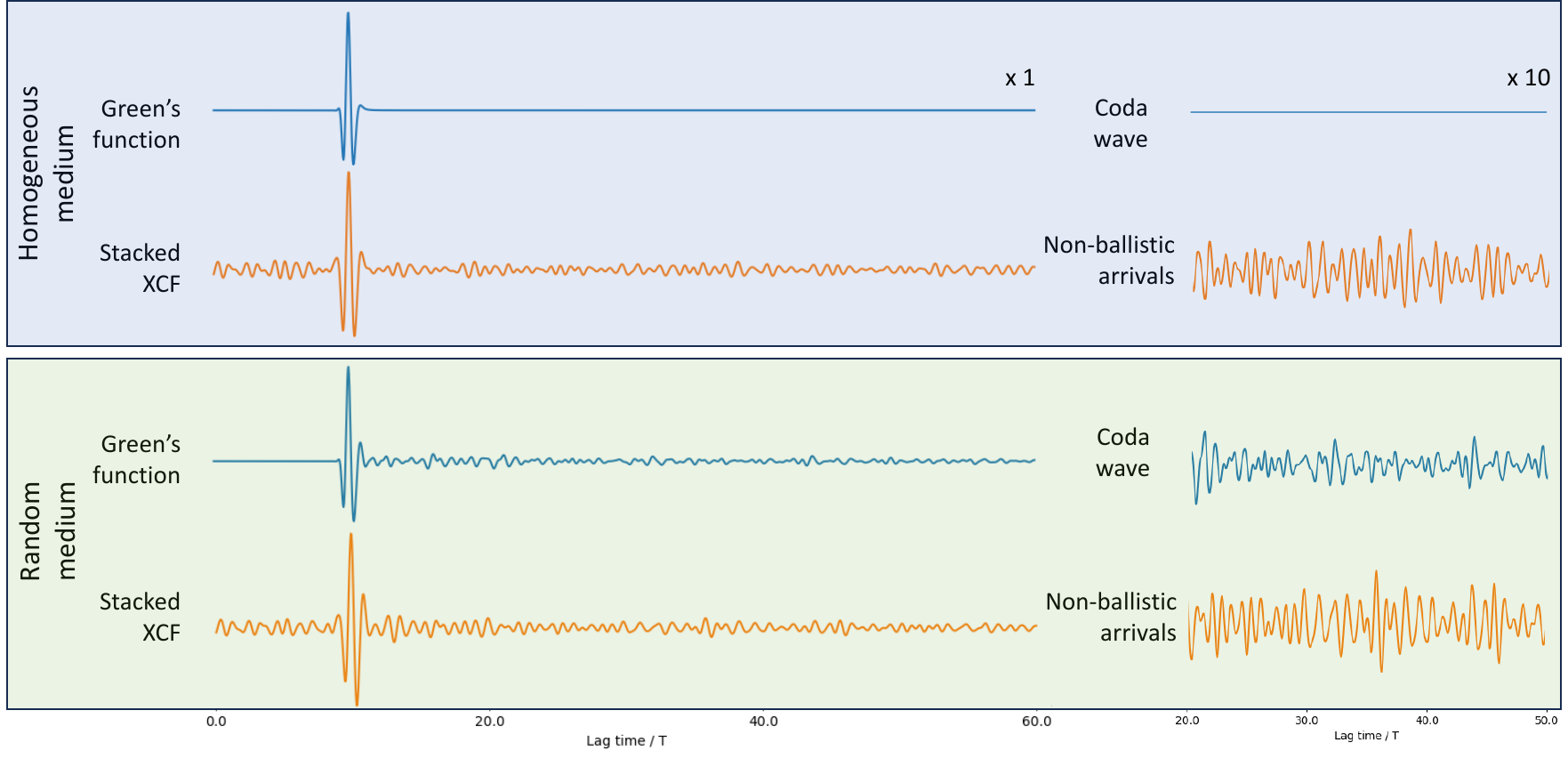}
    \caption{Simulation results comparing the actual Green's function and the stacked cross-correlation function (marked as XCF) in different media. The top panel shows the simulation results in a homogeneous medium. The bottom panel shows the simulation results in a random medium with strong randomly distributed density heterogeneities.}
    \label{fig:GFneqXCF}
\end{figure}

Figure \ref{fig:GFneqXCF} shows a straightforward comparison between the actual Green's function and the stacked cross-correlation function (marked as XCF) in different media. We stack the cross-correlation function over a duration of $180,000$ times the dominant period ($T$) of the ballistic arrival. For 1 Hz ($T = 1$ s) waves, this is equivalent to stacking over two days (48 hours) of noise recordings. The waveforms shown in Green's function and the XCFs are between 0 and $60T$. We take the waveforms between $20T$ and $50T$ for a more detailed comparison. 

In the homogeneous medium, the non-ballistic arrivals in the stacked XCF are generated by the correlations of the recordings from different sources. Due to the random and continuous nature of the noise source environment in the field, these cross-talk artifacts cannot be fully eliminated even if the stacking power is maximized (Figure \ref{fig:stackpower}). As these non-ballistic arrivals mainly reflect the correlation characteristics of the source, they do not deterministically inform the propagation medium (which is free of any scattering in the homogeneous case). 

A remarkable and alarming observation is that we cannot distinguish the non-ballistic arrivals in the stacked XCF simulated in a homogeneous medium from the coda waves simulated in the random medium. This ambiguity challenges existing methods and workflows for non-ballistic arrival interpretation. The XCF non-ballistic arrivals simulated in the random medium are evidently different from the coda waves in the actual Green's function but rather strongly contaminated by the source-induced crosstalk noise.

\begin{figure}[ht]
    \centering
    \includegraphics[width=1\linewidth]{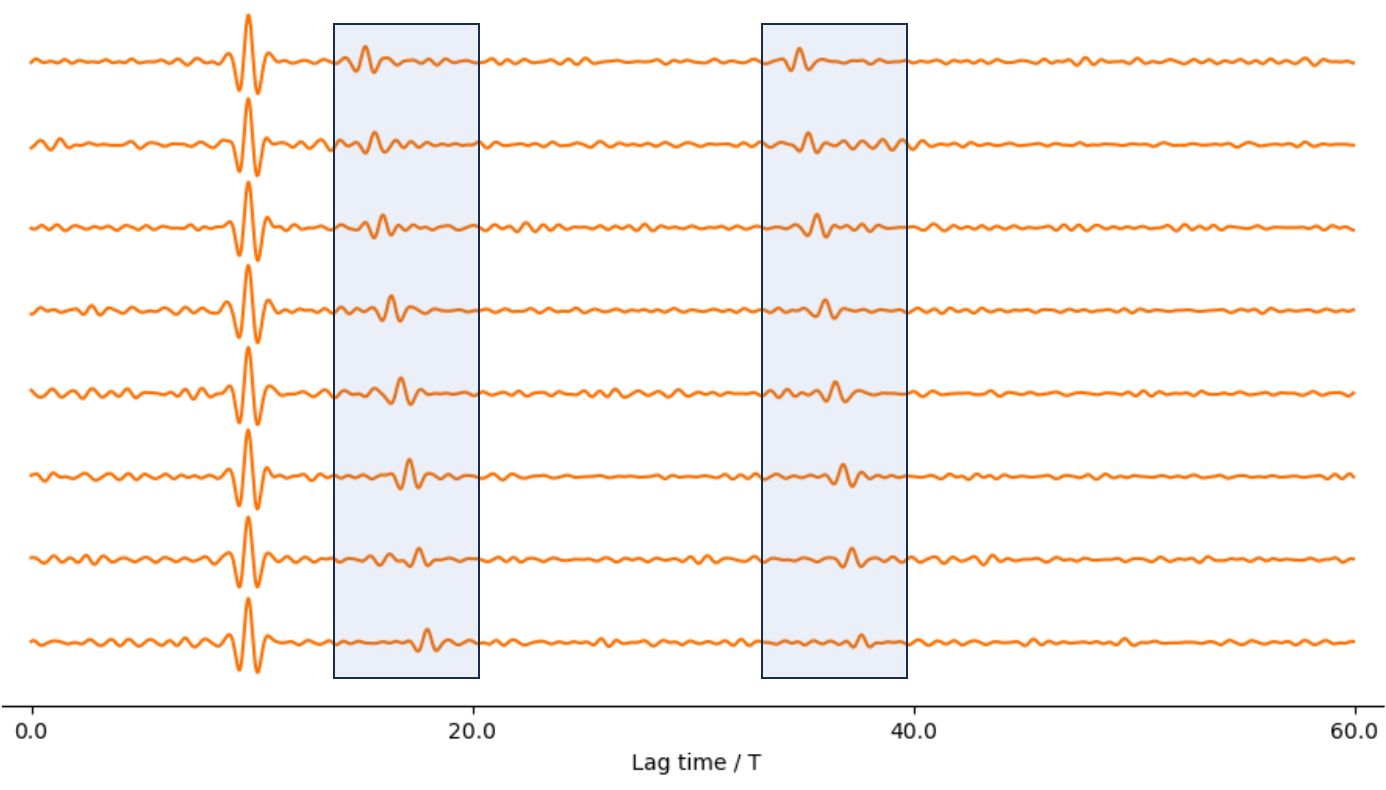}
    \caption{Eight stacked XCFs simulated in a homogeneous medium with correlated noise sources. Blue-shaded zones highlight stable non-ballistic arrivals with phase shifts due to changes in source correlation time. The homogeneous propagation medium is kept constant during different simulations.}
    \label{fig:srcshift}
\end{figure}

Moreover, the theoretical analyses show that the non-ballistic arrival artifacts are stable, as long as the \textit{averaged source spectra, source distribution, and source correlations are statistically stable}. When the noise sources are ocean waves, road traffic, or other fixed location anthropogenic noises, these conditions are often satisfied in the real world. Standard ambient noise processing operations such as amplitude normalization and spectral whitening further enhance the spectra stability. Therefore, the emergence of stable arrivals in the non-ballistic arrivals in the stacked XCF cannot be uniquely interpreted as scattering events in the subsurface. The time shifts measured in these stable non-ballistic arrivals cannot be unequivocally interpreted as changes in the propagation medium without further constraints. 

Figure \ref{fig:srcshift} shows an example where random sources with stable time correlations are simulated in a homogeneous medium. We set the medium parameter constant over time, simulate 8 different realizations of the random source field with the same spectral and location distribution, and only change the source correlation time. This could correspond to gradual changes in ocean wave frequency due to tidal forces or minute changes in traffic speed on the highway. In the resulting cross-correlation functions, the arrival time of the ballistic wave does not change, which properly reflects the constant nature of the propagation medium, while clear time shifts are observed for the stable non-ballistic arrivals in the later time, which is purely due to the shifts in the source correlation time. The relative time shift $dt/t$ is measured around 2\% for the first non-ballistic wave train, and around 1\% for the second wave train. In either case, it would be a mistake if these shifts were interpreted as changes in the velocity in the propagation medium. 

\section{Conclusions}
Through stationary phase analysis, we provide a theoretical framework to quantitatively understand the non-ballistic arrivals in the stacked cross-correlation function. Our main results show that without further constraints about the noise source environment, it is extremely challenging to distinguish the source-induced correlations from the medium-induced correlations in the non-ballistic arrivals. A general equivalency between the later-time arrivals in the stacked cross-correlation functions and coda waves from impulsive sources does not exist. Therefore, interpretation of the non-ballistic arrivals in the seismic ambient noise community requires consideration of all possible scenarios before they are translated into the propagation medium properties deterministically. When the primary noise source correlations are sufficiently ruled out, we provide a theoretical understanding of the stationary phase zones for special cases of the scattering waves. These theoretical results will direct future research to extract more reliable scattering information from the noise correlation functions for higher-resolution imaging and monitoring.

\section*{Open Research Section}
While this is primarily a theory paper and no field data are used, we will organize and upload codes to generate the synthetic data used in this research on GitHub. 

\acknowledgments
The authors acknowledge critical discussions with Jonathan Delph, Arthur Cheng, and Xiaotao Yang during the formation and writing of the research. Y. E. Li acknowledges Jiquan Wang and Cheng-Ju Wu for their help in deriving the stationary phase conditions for planar reflections. Y. E. Li and F. Zhu are supported by the startup grant at Purdue University and the Cheng-Wong Family Charitable Foundation. Jizhong Yang is supported by the National Natural Science Foundation of China under Grant No. 42374136 and the Fundamental Research Funds for the Central Universities of China.

\bibliography{autocorr}

\end{document}